\documentclass[twocolumn,preprintnumbers,amsmath,amssymb]{revtex4-1}
\usepackage{graphicx}
\usepackage{dcolumn}
\usepackage{bm}
\begin{document}

\title{ Efficient Evaluation of Arbitrary Static Fields For Symplectic Particle Tracking}
\author{Lajos Bojt\'ar}
\email{ Lajos.Bojtar@cern.ch }
\affiliation{ CERN, CH-1211 Gen\`eve 23\\}
\date{\today}

\begin{abstract}
This article describes a method devised for efficient evaluation of arbitrary static magnetic and electric fields in a source free region needed for long time tracking of charged particles. Field values given on the boundary of the region of interest are reproduced inside by an arrangement of hypothetical magnetic or electric monopoles surrounding the boundary surface. The vector and scalar potentials are obtained by summing the contributions of each monopole. The second step of the method improves the evaluation speed of the potentials and their derivatives by orders of magnitude. This comprises covering the region of interest by overlapping spheres, then calculating the spherical harmonic expansion of the potentials on each sphere. During tracking, field values are evaluated by calculating the solid harmonics and their derivatives inside a sphere containing the particle. Software has been developed to test and demonstrate the method on a small particle accelerator. To our knowledge, there is no other method of this kind, allowing long time symplectic integration in general static fields, without simplification.
\end{abstract}

\maketitle

\section{ Introduction }
Long time tracking of charged particles in electromagnetic fields is an important topic in several areas of physics, and it is fundamental for particle accelerators. There are thousands of particle accelerators in the world, and their number is growing each year. They are used for various purposes in particle physics,  radiotherapy, industry, as synchrotron light sources, and other applications. Charged particle tracking in accelerators has a long history, and there are lots of software for this purpose. See \cite{0305-4470-39-19-S03} for a historical overview. Despite the clear interest, before this work, there was not an efficient method available for long time particle tracking in general magnetic field in circular accelerators. Existing methods for general magnetic fields are limited to short or moderate-term tracking due to reasons we will discuss later. The usual methods for long time tracking are not general. They always use some simplification or idealization of the field. These simplifications are often justified, especially for rings with a large circumference. There are however many important cases when these simplifications are not valid. This is often the case for small rings or rings containing special magnetic elements.

  A useful method obtains the field values efficiently, in a way that doesn't introduce errors distorting the long time evolution of the simulation.
We describe such a scheme for arbitrary static fields, then confirm it on a small particle accelerator containing eight bending magnets. Although we have not yet implemented the program for static electric fields, we expect it to work for that case and should be easy to include into the present implementation. We do not discuss time-varying fields and fields arising from space-charge effects in this paper.

Errors during tracking of charged particles have two main sources. The first comes from the evaluation of the fields and the second is due to the method of integration. Symplectic integrators are the preferred choice for long term tracking. They preserve phase space volume and keep the energy error introduced by the integration bounded. Although the truncation error of symplectic integrators is not smaller in general compared to conventional integrators, their long term behavior is much better. The energy error of symplectic integrators oscillates around the exact value while using conventional integrators the error accumulates and tends to grow. This is because the dominant term in truncation error is dissipative for conventional integrators and it is non-dissipative for symplectic ones \cite{feng2010symplectic}.

Our study addresses the first source of error, due to the evaluation of the fields.
Symplecticity is a property of the integrator. Whatever bad approximations of
the potentials are given to a symplectic integrator, the integration step will be
symplectic, assuming that the derivatives of the potentials are defined. However, if the magnetic or electric fields described by the potentials do not satisfy certain conditions, the result of the tracking will not correspond to the real dynamics of the system, the potentials will represent something else than one intended to simulate.

Any static magnetic or electric field in a region free of sources has zero divergence, zero curl and the components of the fields are harmonic functions satisfying the Laplace equation.
\begin{align}
\nabla \cdot \textbf{B} &=0 & \nabla \cdot \textbf{E} &= 0 \label{div_cond}\\
\nabla  \times \textbf{B} &= \textbf{0} & \nabla  \times \textbf{E} &=  \textbf{0} \label{curl_cond}\\
\nabla^2 \textbf{B} &= \textbf{0} & \nabla^2 \textbf{E} &= \textbf{0} \label{laplace_cond}
\end{align}
\linebreak
To our knowledge, there is no 3D grid-based interpolation method satisfying all these conditions for general domains. Grid methods, for example \cite{Lekien-MarsdenInt}, don't satisfy the conditions in Eqns.(\ref{div_cond}-\ref{laplace_cond}) between the grid points. Attempts were made to overcome these shortcomings in \cite{DeHaanInt}, but then the $C^1$ continuity of \cite{Lekien-MarsdenInt} is lost between the cells of the 3D grid. 

In the domain of accelerator physics, one approach to deal with the problem of the representation of 3D fields is to take the Fourier transform of the B field in three variables \cite{SLAC_Chang,epac2006_Bahrdt,Titze2016}. Individual magnets can be represented this way but then remains the problem of how to enter and exit from the fields. Our experience showed that cutting the fields even very far from the magnet leads to discontinuities in the field and important energy drift during tracking. We avoid this problem by expanding the entire beam region, which is not a feasible solution with the existing Fourier transform methods. Another problem is that some fields have fast changing localized features. In those cases, the high number of terms required to represent the field make Fourier transform methods impractical. For specific areas like magneto-spheric physics, similar solutions exist \cite{doi:10.1029/2005JA011382}, however, it cannot always be adopted to particle accelerators for the same reasons.

Surface methods describe the fields on the boundary surrounding the region of interest.  Field values on a closed surrounding surface determine the magnetic or electric fields in a source free region. This is true because these fields are harmonic functions satisfying the Laplace equation which has a unique solution for a given boundary condition. A surface method able to express arbitrary static magnetic fields in a way that they satisfy the relevant conditions in Eqns.(\ref{div_cond}-\ref{laplace_cond}) is given in~\cite{Dragt:2010cv}. Since our method is similar to that stage, we give a summary here.

Based on the Helmholtz decomposition theorem, it is possible to describe a general static magnetic field in a source free region as taking the curl of the sum of two vector potentials. One depends only on the tangential component of the $\mathbf{B}$ field  and the second depends only on the normal component. Both are given on the boundary surface surrounding the region of interest.
\begin{equation}
\mathbf{A} = \mathbf{A_t}+\mathbf{A_n} \label{eq_asum}
\end{equation} 
\begin{equation}
\mathbf{A_t}  = -\frac 1 {4\pi} \int_{S'} \mathbf{n(r')}\times\frac{\mathbf{B} (\mathbf{r}')}
{|\mathbf{r}-\mathbf{r}'|}  \mathrm{d}S'  \label{eq_at}
\end{equation}
\begin{equation}
 \mathbf{A_n}  = \int_{S'} [\mathbf{n(r')}\cdot {\mathbf{B} (\mathbf{r}')}]
\mathbf{G^n}[\mathbf{r};\mathbf{r}', \mathbf{m} ]  \mathrm{d}S' , \label{eq_an}
\end{equation}
where
\begin{equation}
 \mathbf{G^n(r;r',m)} = \frac{\mathbf{m} \times \mathbf{(r-r')} }{4 \pi[\mathbf{|r-r'|-m \cdot (r - r')}] \mathbf{|r-r'|} } \label{eq_Am}
\end{equation}
is the vector potential of the Dirac monopole with unit charge, $\mathbf{r'}$ is the location of the monopole, $\mathbf{r}$ is the evaluation point and $\mathbf{m}$ is a unit vector pointing toward the direction of the Dirac string. To get $\mathbf{A_n}$ and  $\mathbf{A_t}$ the bounding surface is partitioned to non-overlapping coordinate patches, then  Eqns.(\ref{eq_at},\ref{eq_an}) are integrated by high order cubature formulas for each patch and summed together. The vector potential is related to the $\mathbf{B}$ field as:
\begin{equation}
\mathbf{B} = \nabla\times \mathbf{A}. 
\end{equation}

\section{ Discussion }
\subsection{Solid harmonics and Taylor expansion } \label{subsection_exp}
Direct use of this surface method would require integration on the boundary surface for each element in the accelerator at each time step. In theory, one can use this method directly for long time tracking together with a symplectic integrator, in practice this would be extremely slow. We haven't found any paper describing a direct usage of the method, most probably due to performance reasons.

A more localized description of the fields is needed to speed up the evaluation. One possibility is to expand the potentials into Taylor series around some reference trajectory. That is the approach taken in \cite{Dragt:2010cv} to construct transfer maps for beamline elements. Particle tracking with transfer maps is fast and useful in beamlines. This approach, however, is not suitable for long time tracking as a remark suggests in A.J. Dragt's vast online book \cite{DragtBook2018} on page 27: "In particular, in the context of Accelerator Physics, the long-term goal of map methods is to be able to describe, predict, and control nonlinear properties with the same facility with which we now handle linear properties. Much has been accomplished in this direction, particularly with regard to single-pass systems and short-to-moderate-term behavior in circulating systems."

The problem is how to enter and exit the field of a magnet. As we mentioned earlier, any cut of the fields leads to a non-negligible discontinuity, which causes an energy drift during tracking. Our method avoids this problem by not having cuts at all. This is not possible with transfer maps, because they must start and end somewhere. A sizeable part of the book \cite{DragtBook2018} is dedicated to mitigating this problem on a case-by-case basis.

One might imagine an expansion of the potentials in terms of Taylor series in the entire volume of interest in 3 spacial variables to avoid the problem of field cuts. This would lead to a method enormously inferior compared to ours for reasons explained below.

Spherical harmonics form a complete set of orthogonal functions and can approximate functions on a sphere to any precision \cite{SHUnitSphere,axler2001harmonic}. The components of the vector potential and the scalar potential are harmonic functions. Inside a sphere, they are uniquely determined by their values on the sphere, therefore spherical harmonics approximations of the potentials on a sphere can also approximate them inside. Spherical harmonics scaled appropriately are called solid harmonics. Regular solid harmonics are the canonical representation for harmonic functions inside a sphere \cite{SHUnitSphere}.

It is important to understand the differences between multipole expansions in terms of Taylor series and terms of solid harmonics, see \cite{multipoleComp} for a detailed comparison. Multipole expansion of potentials with Taylor series has $(\ell +1)(\ell+2)/2$ coefficients for each degree $\ell$. The solid harmonics expansion has only $(2 \, \ell +1)$. That makes a huge difference in the number of calculations. The Taylor expansion needs the partial derivatives of the potentials in 3 variables while the solid harmonic expansion is performed by integration on a sphere only in 2 variables.  Taylor expansion can represent any function in a volume, but solid harmonics can represent only harmonics functions, exactly what we need.

To approximate the fields in some general volume, we need to cover the volume by overlapping spheres. Any location where a particle can go during the tracking should be in a sphere. The potentials inside each sphere are continuous and satisfy the conditions given in Eqns.(\ref{div_cond}-\ref{laplace_cond}). However, when the evaluation algorithm has to switch between the spheres at some step, there is an apparent discontinuity of the potentials and their derivatives between the spheres.
\subsection{The effect of discontinuities }
As we mentioned in the introduction, symplecticity is a property of the integrator, and it is not affected by the quality of the approximation of the potentials if the potentials and their derivatives are defined everywhere, which is the case. Any point inside the volume of interest belongs to the interior of a sphere covering the volume and for any point inside a sphere the potentials and their derivatives can be calculated analytically with the formulas given later. The effect of discontinuities is an energy drift during the tracking. The drift is caused by the fact that the approximations of the potentials inside the spheres are exact only when the potentials don't contain higher other components than the maximum degree of the solid harmonics expansion.
As we limit the solid harmonics expansion at some degree $\ell_{max}$ there is a
small difference between the exact value of the potentials and the approximated
ones. In the overlapping region of two neighboring spheres the approximated
values are slightly different, depending on which sphere was chosen. 

\begin{figure}
\includegraphics [scale=0.7]{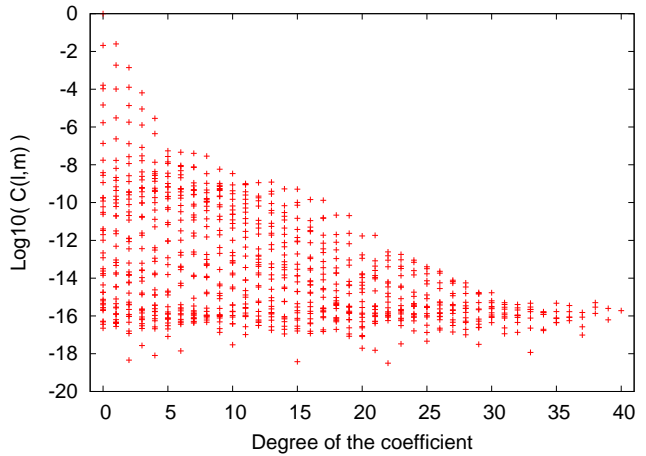}
\caption{\label{fig:clm40} Spherical harmonics coefficients $c_{\ell,m}$ arising during the approximation of one component of the vector potential inside a sphere placed into the end field of a bending dipole magnet. The coefficients are plotted on a base 10 logarithmic vertical scale and normalized to one. The horizontal scale signifies the degree $\ell$ of the spherical harmonics coefficients. Each dot corresponds to a particular order ($m$ in $c_{\ell,m}$ ). We plotted only the positive orders, for this particular example the coefficients belonging to negative orders are zero.}
\end{figure}

 It is possible to reduce this discontinuity as small as needed within the limitations imposed by the numerical precision of the implementation. The higher the degree of the solid harmonics expansion, the smaller are the discontinuities. That is true by assuming that the amplitudes of the coefficients are decreasing with higher degrees. This is usually the case for particle accelerator magnets. FIG.\ref{fig:clm40} shows an exponential decay of the coefficients for a bending dipole magnet as an example. The exponential decrease of the coefficient amplitudes is also a property of other types of magnets, like quadrupoles, octopoles, solenoids, and other kinds of magnets. This example shows that the coefficients with a degree above 40, contribute little. They are around $10^{-16}$, the numerical resolution of 64-bit floating-point numbers.
 
One might ask what the difference between the discontinuities in our method and De Haan's approach described in \cite{DeHaanInt}, which we discarded for long time tracking due to the discontinuities is. Both methods must keep the discontinuities at a very low level. When the potentials are approximated by solid harmonics expansion, the error of the approximation is about the same order of magnitude as the solid harmonics with degree $l_{max}+1$ in the expansion. As FIG.\ref{fig:clm40} shows, the coefficients decrease very fast, exponentially. Therefore the discontinuities between the spheres also decrease exponentially as we go higher with the degree of expansion, $l_{max}$.

In De Haan's 3D grid method the discontinuities can be reduced by decreasing the distance between the grid points. The number of grid points needed to reduce the distance by some factor goes up with the third power of that factor. The error is also decreasing with the third power of the distance between grid points. Additionally, when the grid size is decreased, the number of discontinuities per unit length is increased. Putting all this together we have a slower than linear improvement of the discontinuity errors as the number of grid points is increased. Compare this to the exponential decrease of the spherical harmonics expansion. To keep the effect of the discontinuities at an acceptable level, the number of grid points needed for De Haan's method is too big to be practical. 

When approximating harmonic functions, any 3D grid approach is inferior to ours because harmonics functions are uniquely determined by the boundary conditions. In a 3D grid, only the points on the surface of the volume contain useful information, all the grid points inside the volume are redundant.

\subsection{Evaluation of the potentials} \label{subsection_pot_ev}
 The value of the potentials can be evaluated inside a sphere  with radius R with the following expression. In fact, this is the well known formula for regular real solid harmonics inside a unit ball, scaled by $R$. 
\begin{equation}
f(r,\theta,\phi) = \frac{1}{R} \, \sum_{\ell=0}^{\ell _{max}} \, \sum_{m=-\ell}^{\ell}\, r^\ell  \, c_{\ell m} \, \tilde{P}_\ell ^m (\cos\theta) \, \Phi(\phi;m)  \label{eq_pot_value}
\end{equation} 
We denoted the $\phi$ dependent part as $\Phi(\phi;m)$ given by
\begin{equation}
\Phi(\phi;m) =
\begin{cases}
 \displaystyle \sqrt{2} \sin (|m|\phi)  &\mbox{if } m<0 \\
 \displaystyle 1 &\mbox{if } m=0\\
 \displaystyle \sqrt{2} \cos (m \phi) &\mbox{if } m>0
\end{cases}. \label{eq_Phi}
\end{equation}
In the expression above, $\tilde{P}_\ell ^m (\cos \theta)$ is the orthonormalized associated Legendre polynomial, given as:
\begin{equation}
\tilde{P}_\ell ^m (\cos \theta) = \sqrt{{(2\ell+1)\over 4\pi}{(\ell-|m|)!\over (\ell+|m|)!}}  \, P_\ell^m ( \cos{\theta} ) . 
\end{equation}
$P_\ell ^m (\cos \theta)$ is calculated without the Condon-Shortley phase. In Eq.(\ref{eq_pot_value}),  $ r = |\mathbf{r_e-r_c}|/R $ is the scaled distance between the sphere center $\mathbf{r_c}$ and the evaluation point $\mathbf{r_e}.$  

In the actual implementation, the order of the summation is exchanged as \cite{NumRec} recommends it. Also, the multiplication with the coefficients $c_{\ell m}$ for the negative and positive $m$ values are calculated at the same step. This is the usual technique in most optimized codes \cite{2014arXiv1410.1748L}. It allows a reduction of the number of operations in the $m$ loop by a factor of two.

The real-valued coefficients $c_{\ell m} $ are pre-calculated by the following integral:
\begin{equation}
c_{\ell m} =\int_{\mathbb{S}^2}  \varphi(\theta,\phi) \, \tilde{P}_{\ell}^{m} (\cos \theta) \, \Phi( \phi; m) \, d\Omega ,  \label{eq_coeff_calc}
\end{equation}
where $\varphi(\theta,\phi)$ is the potential to be approximated, and $\mathbb{S}^2$ denotes the surface of the sphere. There are several ways to do this integration. A good comparison of different methods can be found in \cite{Beentjes2016}. Our priority is numerical accuracy, and secondly to minimize the number of evaluations of the potentials $\varphi(\theta,\phi)$, because the computational cost of calculating the potentials by integration of Eqns.(\ref{eq_at}, \ref{eq_an}) or by the method we will describe later, is much higher than the cost of calculating
$P_{\ell}^{m} (\cos \theta)$. Usually Eq.(\ref{eq_coeff_calc}) is evaluated by some numerical quadrature. An elegant way is, to sum values $\varphi(\theta,\phi)$ on the surface of the sphere at the t-design points with equal weights. A spherical t-design is a set of N points on the sphere, such that a quadrature with equal weights using these nodes is exact for all spherical polynomials of degree at most t. More on t-designs can be found in \cite{Womersley:t-design}. An advantage of integrating on a sphere with t-designs is that the sample points has a uniform distribution, contrary to the often used Gaussian quadrature, which has a denser sampling around the poles. As  \cite{Lebedev:squadratura} points out, when the sampling points are distributed uniformly on the sphere, the error in the integral is minimized, assuming the errors in the sample values have a normal distribution. 

For a set of N points $\{ \mathbf{x_i} \}$ in a t-design the following expression is exactly true for all spherical polynomials of maximum degree t or below.
\begin{equation}
\int_{\mathbb{S}^2}  \varphi(\mathbf{x}) d\Omega =\frac{4 \pi}{N}\sum_{i=0}^{N-1}\,  \varphi(\mathbf{x_i}). \label{eq_int_by_sum}
\end{equation}

Combining  Eqns.(\ref{eq_coeff_calc}, \ref{eq_int_by_sum}), the spherical harmonics coefficients can be calculated with the following sum:
\begin{equation}
c_{\ell m} = \frac{4 \pi}{N} \, \sum_{i=0}^{N-1}\,  \varphi(\mathbf{x_i}) \, \tilde{P}_\ell ^{m} (\cos \theta(\mathbf{x_i})) \, \Phi( \phi(\mathbf{x_i});m), \label{eq_clm_tdesign}
\end{equation}
where N is the number of quadrature points in the t-design. The spherical coordinates $\theta(\mathbf{x_i})$ and $\phi(\mathbf{x_i})$ are expressed in terms of the vectors  $\mathbf{x_i}$ pointing to the quadrature points. The author of \cite{Womersley:t-design} has published a set of files containing t-designs up to degree 325 on his website for download. A subset of these files has been used in our implementation.
\subsection{Evaluation of the derivatives} \label{subsection_der_eval}

Apart from the values of the potentials, we also need their first derivatives to integrate the equations of motion. The values and the derivatives are calculated in terms of spherical coordinates first, then converted to Cartesian coordinates. Spherical coordinates are the natural choice in a sphere, but we have many of them, so a global coordinate system common to all spheres is needed. The most convenient is to use Cartesian global coordinates.  The first derivatives in the spherical coordinate $r$ is directly obtained from Eq.(\ref{eq_pot_value}) by differentiation as:
\begin{equation}
\frac{\partial f(r,\theta,\phi)}{\partial r} = \frac{1}{R} \, \sum_{\ell=0}^{\ell _{max}} \, \sum_{m=-\ell}^{\ell}\,  \ell  \, r^{\ell-1}  \, c_{\ell m} \, \tilde{P}_\ell ^m (\cos\theta) \, \Phi( \phi;m). \label{eq_r_der}
\end{equation}
Similarly for $\phi$
\begin{equation}
\frac{\partial f(r,\theta,\phi)}{\partial \phi} = \frac{1}{R} \, \sum_{\ell=0}^{\ell _{max}} \, \sum_{m=-\ell}^{\ell} \, r^\ell  \, c_{\ell m} \, \tilde{P}_\ell ^m (\cos \theta) \, \frac{\partial \Phi( \phi;m) }{\partial \phi} . \label{eq_phi_der}
\end{equation}
The derivative of $\Phi(\phi;m)$ must be treated for the three case listed in Eq.(\ref{eq_Phi}).

The derivatives are expressed in terms of orthonormalized associated Legendre polynomial to achieve good numerical precision. As \cite{NumRec} points out, calculating $P_\ell ^m (\cos \theta)$ directly without normalization, produces much more significant numerical errors than the calculation with the orthonormalized version, therefore it is not advised.

We got the $\theta$ derivative of the potentials by the following expression:
\begin{equation}
\frac{\partial f(r,\theta,\phi)}{\partial \theta} = 
\frac{1}{R} \, \sum_{\ell=0}^{\ell _{max}} \, \sum_{m=-\ell}^{\ell} \, r^\ell  \, c_{\ell m}\, \Phi(\phi;m)  \, \frac{\partial \tilde{P}_{\ell} ^m (\cos \theta)}{\partial \theta} , 
\end{equation}
where the $\theta$ derivative of $\tilde{P}_{\ell}^m (\cos \theta)$ is calculated for the case $m \neq 0$ as:
\begin{equation}
\begin{aligned}
\frac{\partial \tilde{P}_{\ell} ^m (\cos \theta)}{\partial \theta} = \frac {k_{\ell m} (\ell+1-m)  \tilde{P}_{\ell+1} ^m (\cos (\theta))}{\sin(\theta)}  \\
-\frac {\cos (\theta) \, (\ell+1)  \tilde{P}_{\ell} ^m (\cos (\theta))}{\sin(\theta)} , \label{eq_theta_der_m_neq_0}
\end{aligned}
\end{equation}
with the pre-calculated constants
\begin{equation}
k_{\ell m}= \sqrt{\frac{(2 l+1) (l- m )!}{( m +l)!}} \, / \sqrt{\frac{(2 l+3) (1+l -m)!}{(m +l+1)!}} .
\end{equation}
In Eq.(\ref{eq_theta_der_m_neq_0}) there is a singularity due to the division by $\sin(\theta)$. This is no singularity in the potentials, only a consequence of the spherical coordinate system. It is eliminated by checking $\theta$ for zero. In that case, we replace zero with a tiny number. The $\theta$ derivative will be huge, but this apparent near singular behavior disappears when the derivatives are converted to Cartesian coordinates.

With this treatment of singularity, Eq.(\ref{eq_theta_der_m_neq_0}) gives a precise value only when $m \neq 0$. For the case $m=0$, we use a different formula given as:
\begin{equation}
\frac{\partial \tilde{P}_{\ell} ^0 (\cos \theta)}{\partial \theta} = 
  - \sqrt{l (l+1)} \, \tilde{P}_{\ell} ^1 (\cos \theta) . \label{eq_theta_der_m_eq_0}
\end{equation}
Eq.(\ref{eq_theta_der_m_eq_0}) follows from the fact that
\begin{equation}
\frac{\partial P_{\ell} ^0 (\cos \theta)}{\partial \theta} = 
 - P_{\ell} ^1 (\cos \theta) . 
\end{equation}
Please note the minus sign above. It appears, because $P_{\ell} ^1 (\cos \theta)$ is calculated without the Condon-Shortley phase.
The factor $\sqrt{l (l+1)}$ comes from the normalization.

The potentials and their derivatives are evaluated in the same loop.
The most demanding part computationally is the calculation of $\tilde{P}_{\ell} ^m (\cos \theta)$ values, but it can be reused in Eqns.(\ref{eq_pot_value},\ref{eq_r_der},\ref{eq_phi_der},\ref{eq_theta_der_m_neq_0},\ref{eq_theta_der_m_eq_0}).

\subsection{An alternative calculation of the potentials} \label{subsection_potcalc}

To use the evaluation described above, the coefficients $c_{\ell m}$ must be pre-calculated with Eq.(\ref{eq_clm_tdesign}) for each sphere covering the volume of interest. This requires the calculation of the exact potentials $\varphi(\mathbf{x_i})$. The magnetic potentials can be calculated in certain cases with Eq.(\ref{eq_asum}), which is the method described in \cite{Dragt:2010cv}, but not always.

Consider a magnet and a vacuum chamber touching the poles of the magnet.
The method in \cite{Dragt:2010cv} calculates the vector potential by integrating the magnetic field on a boundary surface, which is the surface of the vacuum chamber now. Eqns.(\ref{eq_at},\ref{eq_an}) describe a layer of dipoles and monopoles on the surface of the vacuum chamber. We want to cover the entire volume of the chamber
with overlapping spheres to approximate the potentials efficiently. The surface of the vacuum chamber will be inside the spheres and also the monopoles and dipoles. This is a problem because the solid harmonics approximation works only if there are no sources inside the spheres. We can use the method in \cite{Dragt:2010cv} only when there is sufficient space between the vacuum chamber and the iron poles to cover the beam region with the spheres without sources inside.

The solution to this problem we devised is to shift the location of the magnetic monopoles further, toward the exterior of the volume of interest. The shift of the monopoles will change the fields, so the Eqns.(\ref{eq_asum}, \ref{eq_at}, \ref{eq_an}) cannot be used. The strengths of the monopoles have to be changed such, that the field values on the bounding surface stay the same. The monopoles are sufficient to reproduce the field at the bounding surface. No need for dipoles.

\begin{figure}
\includegraphics [scale=0.7]{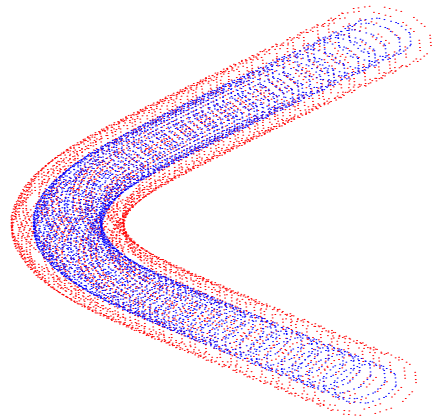}
\caption{\label{fig:bhz} 3D view of monopoles placed around the boundary surface. The red points show the locations of the monopoles and the blue dots are the quadrature points on the bounding surface  where the normal components of the field have to be reproduced by the monopoles.}
\end{figure}

The bounding surface has to be discretized to calculate the strengths of the monopoles. The surface is partitioned into quadrilateral tiles without gaps. Each tile has some number of Gaussian quadrature points. Formulas for placing the quadrature points inside the tiles can be found in \cite{Stroud:104291,FiniteElem}. There are infinite possibilities to put the monopoles around the volume of interest. If the quadrature points of each tile are translated along the normal vector of the bounding surface, then we have a simple method to place the sources some distance away from the boundary. The placement of the monopoles is illustrated in FIG.\ref{fig:bhz} for a bending dipole.

After the  discretization of the bounding surface and the placement of the monopoles we can set up a linear system of $n$ equations to calculate the strengths of $n$ monopoles as: 
\begin{equation}
\mathbf{A.x=b}, \label{eq_matrix}
\end{equation}
where $\mathbf{A}$ is a $n$ by $n$ matrix,  $\mathbf{x}$  and $\mathbf{b}$ are column vectors with $n$ elements. Eq.(\ref{eq_matrix}) is solved for $\mathbf{x}$. The elements of the vector $\mathbf{b}$ are calculated as the dot product of a unit vector normal to the bounding surface at a quadrature point and the magnetic (or electric) field vector at the same point.
\begin{equation}
b_i= \mathbf{\hat{n}_i}.\mathbf{B_i} \label{eq_b}
\end{equation}
The unit vector $\mathbf{\hat{n}_i}$ is normal to the bounding surface at the $i$th quadrature point and it points outward.  Magnetic field values $\mathbf{B_i}$, are given as an input by some magnetic measurement or got from an EM simulation software. The elements $a_{i,j}$ of the matrix $\mathbf{A}$ are the dot products of the normal unit vector $\mathbf{\hat{n}_i}$ and the magnetic (or electric) field vector produced by the $j$th monopole, as:
\begin{equation}
a_{i,j}= \mathbf{\hat{n}_i}. \, \frac{\mathbf{r}_i-\mathbf{r}_j}{
\left\| \mathbf{r}_i-\mathbf{r}_j \right\|^3}, \label{eq_aij}
\end{equation}
where $i,j={1 .. n}$, $\mathbf{r}_i$ is a vector pointing to the location of the $i$th quadrature point and $\mathbf{r}_j$ is a vector pointing to the $j$th monopole. 

Eq.(\ref{eq_matrix}) is solved by LU decomposition, a standard direct method. Once the strengths of the monopoles are obtained, the vector potential can be calculated anywhere inside the volume of interest and to some extent beyond it by summing the contributions from each monopole. The vector potential of the Dirac monopole with unit charge was already given in Eq.(\ref{eq_Am}).

In case of an electric field, the scalar electric potential is $V_{E}(\mathbf{r;r'}) = q/( 4 \pi \epsilon_{0}\left\| \mathbf{r-r'} \right\| )$. 

One must take care of the singularity in Eq.(\ref{eq_Am}), which is along the half line in the direction of the Dirac string. This direction is a free parameter, but it must be set such that it doesn't intersect the beam region. For more about this topic, see  \cite{Dragt:2010cv} and \cite{mitchell2007phd}.

The distance of the monopoles from the bounding surface is also a free parameter. It takes a little experimentation to find the best value. If the monopoles are too close, the result will be less smooth. If they are too far, the solution of Eq.(\ref{eq_matrix}) will be less accurate. There is another consequence of this parameter, which we found rather important. While the magnetic field on the surface stays the same with different distances, the vector potential will be different. The choice of this distance acts as a gauge transformation. This choice has an influence on the amplitude of the energy oscillation during the tracking.
\begin{figure}
\includegraphics [scale=0.7]{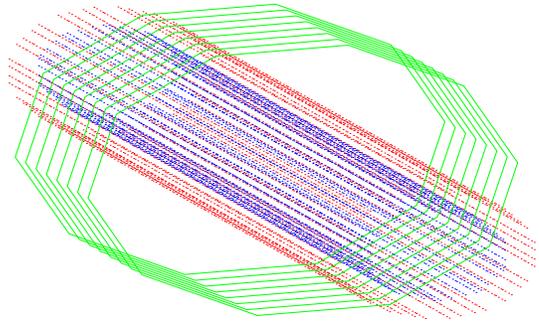}
\caption{\label{fig:solenoid} 3D view of the monopoles and the current loops around the boundary surface of a solenoid magnet. The red points indicate the monopoles, the blue dots the quadrature points on an open cylinder surface and the green polygons are the current loops.}
\end{figure}

The placement of monopoles described above and depicted in FIG. \ref{fig:bhz} works well for all multipoles used in accelerators. In these elements, the longitudinal component of the magnetic or electric field has zero integral around a complete turn of the machine. 

Let's consider now the field of a solenoid magnet. In this case, the integral of the longitudinal component of the magnetic field is not zero along a complete turn. One can describe the $\mathbf{B}$ field of a magnetic monopole not only with the vector potential in Eq.(\ref{eq_Am}) but also with a magnetic scalar potential. Therefore the integral of the longitudinal component of the $\mathbf{B}$ field produced by magnetic monopoles along a complete turn of the accelerator must be zero, because $ \oint \mathbf{ \nabla \varphi(\mathbf{r})} \cdot d{\mathbf{r}} = 0$ for any potential $\varphi(\mathbf{r})$. We can conclude that an arrangement of magnetic monopoles alone cannot reproduce the magnetic field of a solenoid magnet. The longitudinal component of the field has to be supplied by electric currents. That can be achieved by adding one or more current loops around the bounding surface as  FIG. \ref{fig:solenoid} shows. The geometry of the current loops is not important. However, they should not be too close or too far from the bounding surface. They must give the same integral of the longitudinal component of the $\mathbf{B}$ field as the solenoid we want to reproduce. The vector potential is obtained as a sum of the vector potentials of the magnetic monopoles and the current loops.

An arrangement of magnetic monopoles and current loops can describe any static magnetic field. This follows from the Helmholtz decomposition theorem. It states that any smooth rapidly decaying vector field can be described in 3 dimensions as a sum of a curl-free and a divergence-free vector fields. The field to be reproduced is the $\mathbf{B}$ field, the divergence-free field corresponds to the curl of the magnetic vector potential of the current loops and the curl-free field can be associated with the gradient of the magnetic scalar potential of the monopoles. We do not use the magnetic scalar potentials of the monopoles, only the vector potentials, but the $\mathbf{B}$ field obtained from them is the same apart from the singularity at the Dirac string, which is outside of the beam region. 

An arrangement of electric monopoles alone is sufficient to describe any static electric field.

In particle accelerators, an arrangement of monopoles and currents loops depicted on 
FIG. \ref{fig:bhz}. and FIG. \ref{fig:solenoid}. cover most cases.
To make the method complete, we consider a simply connected domain with a closed boundary surface around it. In this case, the static magnetic or electric field lines can not have closed loops inside the volume of interest which is free of sources. Therefore a scalar potential is enough to describe the fields. The $\mathbf{B}$ or $\mathbf{E}$ fields can be reproduced by magnetic or electric monopoles placed outside a closed boundary surface. There is no need for current loops.

Compared to the surface method described in \cite{Dragt:2010cv}, our method is always applicable to calculate the potentials for the solid harmonics expansion, which is not always the case for \cite{Dragt:2010cv} as we explained earlier. When there is enough space to cover the beam region with spheres without having iron inside the spheres, one is free to choose between the two methods. Ours will give a smoother potential for the same number of quadrature points because the monopoles are located further from the bounding surface. With the method in \cite{Dragt:2010cv}, it is easier to increase the number of quadrature points on the bounding surface at the cost of  slower evaluation of $\varphi(\mathbf{x_i})$ during the calculation of coefficients $c_{\ell m}$ in Eq.(\ref{eq_clm_tdesign}), because there is no need to solve Eq.(\ref{eq_matrix}).

\begin{figure}
\includegraphics [scale=0.7]{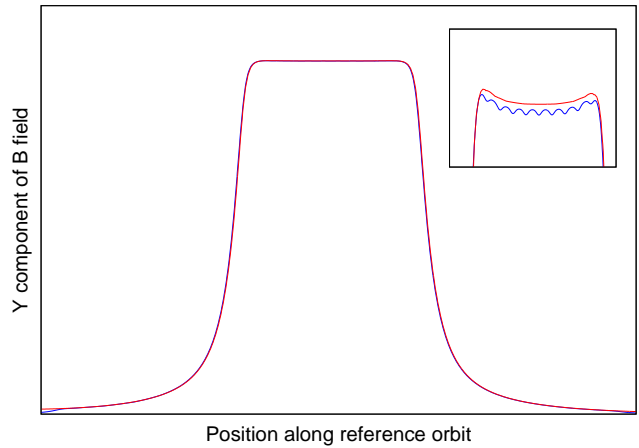}
\caption{\label{fig:bhz_field} The vertical component of the $\mathbf{B}$ field for a sector dipole magnet halfway between the reference orbit and the maximum aperture. The red curve is the reference field calculated by electromagnetic CAD software, and the blue curve is the curl of the vector potential. On the small sub-picture, the Y-axis is zoomed in by a factor 166.}
\end{figure}

FIG. \ref{fig:bhz_field} shows an example for a sector dipole magnet, where the $\mathbf{B}$ was reconstructed as the curl of $\mathbf{A}$ calculated with the procedure above. In this example, there was no need to use current loops, magnetic monopoles around an open cylinder as boundary surface suffices to obtain a good approximation of the $\mathbf{A}$ field.

The accuracy of the reconstructed $\mathbf{B}$ field depends on several factors, most importantly the number of monopoles. In this example, we used 4096 monopoles.  The relative RMS error between the reference $\mathbf{B}$ and $\nabla\times \mathbf{A}$ is $4 \times 10^{-3}$. The errors were a bit smaller for the quadrupole and for the solenoid magnet we tested. In an actual particle accelerator model, one should use a scaled version of $\mathbf{A}$, such that the integrated curl of $\mathbf{A}$ along the reference trajectory matches the magnetic measurement of the accelerator component. After this scaling, a deviation would remain, as illustrated in the magnified part of FIG. \ref{fig:bhz_field}. The undulation of the blue trace is due to the finite number of monopoles. One can reduce it by increasing the number of monopoles. The amplitude of the undulation also depends on the distance from the reference orbit. The farther the evaluation point is from the reference orbit, the bigger the undulation. However, we haven't found any noticeable effect of this on the beam dynamics during tracking.  

\subsection{Preparation of the global field map}  \label{subsec_prep}
To test the method, we assembled a small accelerator from eight sector bending magnets. The edge focusing is sufficient to achieve stable optics when the distances between the dipoles are correctly chosen. The workflow to prepare this simple circular accelerator model comprised of the following steps. Each item in the list corresponds to the same task or execution of some command of the software written. Several of these steps have previously been discussed.
\begin{enumerate}
\item Obtain the reference orbit through the magnet, including most of the fringe fields. This is usually a tracking result in some EM CAD software such as Opera, CST EM Studio, in the form of a text file containing the coordinates of a reference particle trajectory.
\item Construct a boundary surface around the reference orbit by extruding a profile around it, a circle, ellipse or a rectangle in most cases. Partition this surface into quadrilateral tiles each containing a number of Gaussian quadrature points. Elevate the quadrature points from the boundary surface to some distance. Place monopoles, magnetic ones in this example, to the location of the elevated points and write their coordinates into a text file.
\item Get the field values at those coordinates from the EM simulation software of your choice or some magnetic measurements.
\item Calculate the strengths of the monopoles with the field values mentioned in the previous item by solving Eq.(\ref{eq_matrix}). Save the positions of monopoles and their strengths into a text file.
\item Read the file as mentioned above, then translate and rotate the monopoles (and current loops if needed) to the correct location and orientation for each magnet of the accelerator.
\item Set the directions of Dirac strings of the magnetic monopoles such that they don't cross the beam region. That is to avoid problems related to the singularity of the vector potential present in Eq.(\ref{eq_Am}).
\item Cover the entire beam region with overlapping spheres. Then, for each sphere calculate the solid harmonics expansion coefficients with Eq.(\ref{eq_clm_tdesign}) and write these coefficients along with all supplementary information into a binary file.
\end{enumerate}
 
After completing this procedure, a binary file is available containing solid harmonics representation of the vector potential for any point inside the beam region. This field map can be used to evaluate the vector potential and the derivatives as it is described in sections \ref{subsection_pot_ev},  \ref{subsection_der_eval}. Our simple test machine contained only eight instances of the same type of magnetic element. In a real-life scenario, there are many different types of magnets. One would make a global field map for each magnet family driven by the same current circuit. Before simulating with some current setting for the circuits, the spherical harmonics coefficients in each field map must be scaled and summed. This is a rather fast operation. The spherical harmonics expansions of the field maps have to be performed on the same set of spheres covering the beam region, otherwise adding the coefficients together would not make sense. Preparing a global field map this way allows to change current settings for different tracking runs quickly without the recalculation of field maps. The machine in our test contained only magnetic elements, but there is no obstacle to mix static magnetic and electric elements to produce field maps.
\subsection{Symplectic tracking results}
Once a global field map is prepared as described above, containing spherical harmonics expansions of the vector and maybe the scalar potentials everywhere in the beam region, symplectic tracking can be performed. The motion of a charged particle calculated by the usual relativistic Hamiltonian
\begin{equation}
H= \sqrt{m^2 c^4 +c^2( \mathbf{p}-q \mathbf{A} )^2 }+q V_{E} \label{eq:hamiltonian},
\end{equation}
where $\mathbf{A}$ is the vector potential, $V_{E}$ is the electric scalar potential, $\mathbf{p}$ is the canonical momentum and $q$ is the charge of the particle.
The equations of motion are six first order ODE's. Integrating these equations in the long term with classical methods like the Runge-Kutta introduces errors, like the drift of energy or non-preservation of phase-space volume. Symplectic integration methods don't suffer from these problems. 
 Until recently, the general belief was that higher order symplectic integrators for general non-separable Hamiltonians have to be implicit, although for specific cases explicit integrators were available \cite{0305-4470-39-19-S03,PhysRevE.68.046502}.  An explicit second-order method for general Hamiltonians, symplectic in extended phase space, is described in \citep{PhysRevE.94.043303}. We implemented this second-order explicit symplectic integrator, which has the correct long time behavior.
\begin{figure}
\includegraphics [scale=0.7]{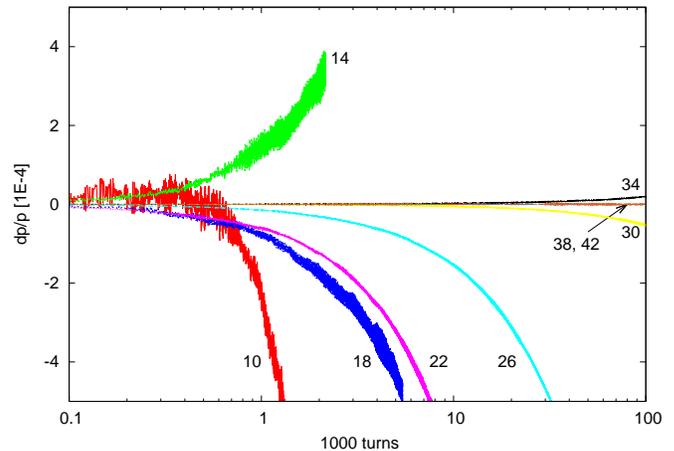}
\caption{\label{fig:momdrift} Drift of momentum during long time tracking using solid harmonics expansions of the potentials with maximum degree $\ell_{max}$ between 10 and 42.}
\end{figure}
As mentioned earlier, the potentials have discontinuities between the spheres. These appear as fictional sources, and they produce an energy drift during the tracking. This energy drift is a distinct phenomenon from the energy drift caused by a non-symplectic integration.

We performed single particle trackings with several $\ell_{max}$ ($\ell_{max}$ is the maximum degree of the solid harmonics expansions) to see the effect of discontinuities. Instead of the energy drift, in FIG. \ref{fig:momdrift} we indicated the relative momentum deviation, more usual in accelerator physics. It can be seen that the drift is decreasing exponentially with the increase of $\ell_{max}$. Above a certain value of $\ell_{max}$, there is no reduction in the drift. Instead, it increases. This is due to the finite precision of the floating point numbers. With the standard double precision, we got the smallest drift about $ dp/p =5 \times 10^{-8}$ in $10^5$ turns with $\ell_{max}=42$. This is good enough for nearly all simulations in accelerator physics. One can make the drift as small as needed, at the cost of more computation by using higher precision floating numbers. 
\begin{figure}
\includegraphics [scale=0.7]{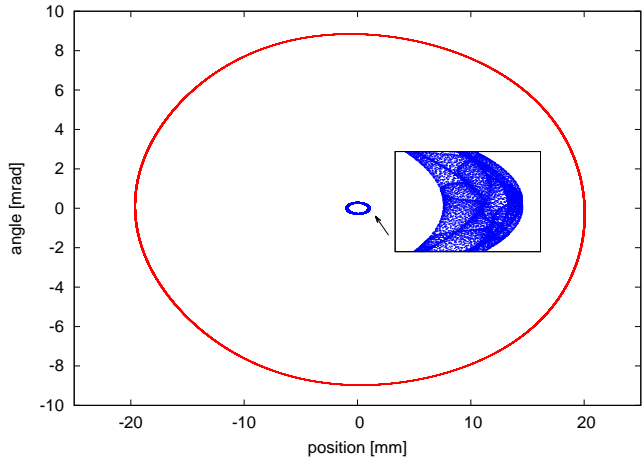}
\caption{\label{fig:phs} Phase space plot for $10^5$ machine turns,  tracking with $l_{max}=38$ spherical harmonics expansion of the vector potential. The red and blue ellipses belong to the horizontal and vertical planes respectively. The amplitude of the vertical motion is smaller than the horizontal to show the interesting effect in the zoomed image. }
\end{figure}

FIG.\ref{fig:phs} shows the phase space plot in the horizontal and vertical planes at a position outside the dipoles field. As expected, with some reasonable tunes of the machine, the phase space points are on a thin ellipse. There is no sign of change in the phase space volume. The effect of the energy drift is negligible, and it is not noticeable on the phase space plot. The phase space points of the vertical plane are on a thicker ellipse, this is the effect of the bigger amplitude motion in the horizontal plane and not due to some artificial phase space volume drift.

\begin{figure}
\includegraphics [scale=0.7]{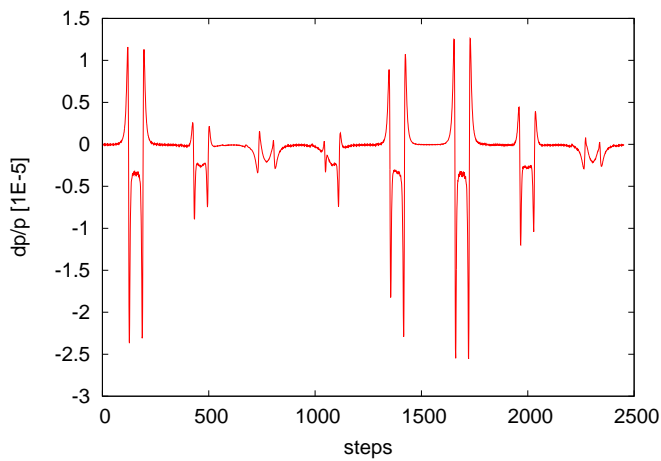}
\caption{\label{fig:dpop} Momentum error arising during tracking a single turn with 1 cm step size.}
\end{figure}
FIG.\ref{fig:dpop} shows the momentum deviation for a single turn around the machine. The long-term drift is not visible in a single turn. The time scale is too short for that. The highest deviation occurs near the edges of the magnets where the vector potential changes fast. One can reduce this by using smaller time steps or using higher order integrator. The second order integrator we implemented needs four evaluations of the potentials and their derivatives at each time step, which in practice is only three because the first evaluation of the current step and last evaluation of the previous step is the same. Yoshida's method \cite{YOSHIDA1990262} allows the construction of higher order integrators, but these require many more evaluations. For the typical relative momentum deviation required for particle accelerators, the second order method might be faster.
\subsection{Implementation and performance}
We implemented the algorithms in this paper and many additional codes for testing different ideas as a Java library named SIMPA. It stands for \underline{s}ymplectic \underline{i}ntegration through \underline{m}ono\underline{p}ole \underline{a}rrangements. This code is not yet a finished tool, and it is not ready for release. More work is necessary to turn it into a tracking tool. It is available however from the author on request by e-mail for people interested in experimenting with it. We plan to release it later as open source code if there is interest.
Although the ideas in this paper were tested in a particle accelerator, we think they are general enough to be useful in other areas of physics as well.

The performance of the method depends on many factors, and there are several computation intensive tasks mentioned in this paper. Some of these are done only once or a few times as part of the workflow listed in section \ref{subsec_prep}. The most important part from the performance point of view is the evaluation of potentials and their derivatives. As one can see in Eqns.(\ref{eq_pot_value}, \ref{eq_r_der}, \ref{eq_phi_der}, \ref{eq_theta_der_m_eq_0}, \ref{eq_theta_der_m_neq_0}) the number of operations is proportional to the square of  $\ell_{max}$, the maximum degree of the solid harmonics expansion. In our test on Intel i7-6700 CPU at 3.40GHz, using a single core with hyper-threading enabled, we got about 58000 evaluations per second of the vector potential and its derivatives with  $\ell_{max}=38$. This is about 16 turns per second for a single particle with 2 cm step size in our test machine. The speed can be improved many times by implementing a multi-threaded version of the evaluation. 

The spherical harmonics expansion of the entire beam region, which comprises 464 overlapping spheres in our test machine, took 880 seconds using all the eight threads. This task is done only once for each magnet group, connected to the same current circuit.

Solving the linear system in Eq.(\ref{eq_matrix}) with 4960 variables took 28 seconds. This task has to be done only in the preparation phase of the model, maybe a few times for each magnet type. 
\section{Concluding remarks}
We developed a method for symplectic long time tracking of charged particles in static magnetic and electric fields. The main idea is to cover the beam region with overlapping spheres and expand the vector and scalar potentials in terms of solid harmonics for each sphere. The discontinuities between neighboring spheres can be reduced as low as needed to perform long time trackings by increasing the degree of the expansion. Such an expansion of the potentials requires the values of the potentials outside the beam region. We devised and described a surface method in section \ref{subsection_potcalc} which allows getting the potentials in an analytic way anywhere in the region of interest and to some extent beyond that. This surface method is general. It can describe any combination of static magnetic and electric fields in a source free region.

These two ideas combined allowed us to demonstrate long time symplectic tracking in a circular accelerator with general static fields for the first time.

One should not mislead by the simplicity of the ring used for the demonstration. The method has been already used for modeling a real machine called ELENA \cite{ElenaDesignRep}, taking full advantage of the unique abilities of our approach. For the first time, we managed to model the effect of toroid kicks due to the magnetic system of the electron cooler in long time trackings. The results of these simulations will be the subject of a different paper.

Our method doesn't aim to compete with established tracking codes like MAD-X. It is somewhat complementary to those. Most tracking codes use idealized elements, often they are "kick-codes" and therefore much faster and easier to use. The algorithm described here is useful when those codes can not be used, or their model is not accurate enough due to the idealization of accelerator components. This can be the case for small machines with significant fringe fields or machines with special elements. One example of such a machine is ELENA, where the magnetic field of the electron cooler has a considerable influence on the beam dynamics. Despite a significant effort, previous attempts failed to perform long time tracking in ELENA including important effects due to the toroid kicks of the electron cooler \cite{Priv_Comm_Beloshit}.

We implemented software to demonstrate and test the algorithm. If there is sufficient interest, our software can be turned into a general purpose library for evaluating static magnetic or electric fields or into a more specific tool for symplectic tracking in accelerators.
\begin{acknowledgments}
I would like to express my gratitude towards Andrea Latina from CERN BE-ABP and Pavel Belochitskii from CERN BE-OP for their valuable comments and encouragements.
\end{acknowledgments}

\bibliography{simpa} 

\end{document}